\documentclass[%
 aip,
 amsmath,amssymb,
 reprint,%
]{revtex4-1}

\usepackage{graphicx}% Include figure files
\usepackage{dcolumn}% Align table columns on decimal point
\usepackage{bm}% bold math
\usepackage{epstopdf}

\usepackage[utf8]{inputenc}
\usepackage[T1]{fontenc}
\usepackage{mathptmx}

\usepackage{textcomp}

\begin{document}

\preprint{AIP/123-QED}

\title[Rotational state-changing collisions of C$_2$H$^-$ and C$_2$N$^-$ anions with He]{Rotational state-changing collisions 
of C$_2$H$^-$ and C$_2$N$^-$ anions with He under interstellar and cold ion trap conditions: a computational comparison.}
% Force line breaks with \\

\author{Jan Franz}
\affiliation{ 
Department of Theoretical Physics and Quantum Informatics, Faculty of Applied Physics and Mathematics, Gda\'nsk University of 
Technology, ul. Narutowicza 11/12, 80-233 Gda\'nsk, Poland
}
\author{Barry Mant}
\affiliation{ 
Institute for Ion Physics and Applied Physics, University of Innsbruck, Technikerstr. 25/3, 6020 Innsbruck, Austria
}
\author{Lola Gonz\'alez-S\'anchez}
\affiliation{ 
Departamento de Qu\'imica F\'isica, University of Salamanca, Plaza de los Ca\'idos sn, 37008 Salamanca, Spain
}
\author{Roland Wester}
\affiliation{ 
Institute for Ion Physics and Applied Physics, University of Innsbruck, Technikerstr. 25/3, 6020 Innsbruck, Austria
}
\author{Franco A. Gianturco}
\email{francesco.gianturco@uibk.ac.at}
\affiliation{ 
Institute for Ion Physics and Applied Physics, University of Innsbruck, Technikerstr. 25/3, 6020 Innsbruck, Austria
}

\date{\today}% It is always \today, today,

\begin{abstract}

We present an extensive range of quantum calculations for the state-changing rotational dynamics involving two simple molecular anions
which are expected to play some role in evolutionary analysis of chemical networks in the Interstellar environments, C$_2$H$^-$($X^1\Sigma^+$) 
and C$_2$N$^-$ ($X^3 \Sigma^-$) but for which inelastic rates are only known  for C$_2$H$^-$.
The same systems are also of direct interest in modelling selective photo-detachment (PD) experiments in cold ion traps where the He atoms
function as the chief buffer gas at the low trap temperatures.
This study employs accurate, \textit{ab initio} calculations of the interaction potential energy surfaces (PESs) for these anions, 
treated as Rigid Rotors (RR) and the He atom to obtain a wide range of state-changing quantum cross sections and rates at temperatures 
up to about 100 K. The results are analysed and compared for the two systems,  to show differences and similarities between their rates of  state-changing dynamics.

\end{abstract}

\maketitle

\section{\label{sec:introduction}Introduction}

The discovery of carbon chain anions in interstellar and circumstellar media has triggered and stimulated a large number of theoretical and
experimental studies on these species (e.g. see references \citenum{08CoMiWa,09CoMixx.cnm,17MiWaFi.LM}). 
Their structures and spectral features, as well as the clarification of 
their  importance and of their role in interstellar chemistry, and in gas phase ion-molecule reactions in general, have therefore also
attracted many specific studies \cite{08HrOsxx,09WaHaHe,07EiSnBa} on their behaviour. Recent examples have been our experimental work on the absolute photodetachment cross-section measurements for hydrocarbon chain anions  \cite{11BeOtTrHp} which has been  nicely matched and confirmed by a computational study of the same systems  \cite{14DoKoOr}.

The possible, and likely, existence of anions in astrophysical sources was first predicted theoretically and considered in earlier chemical
models \cite{73DaMcxx,81Hexxxx} although the first negative hydrocarbon C$_6$H$^-$ was only detected in 2006 \cite{06McGoGu}, 
thereby also  solving the problem of the unidentified lines discovered by Kawaguchi \textit{et al.} \cite{95KeYaIs}. 
That identification was soon followed by the detection of other negatively charged species such as C$_4$H$^-$ \cite{07CeGuAg}, 
C$_8$H$^-$ \cite{07ReHoLo,07BrGuGo}, C$_3$N$^-$ \cite{08ThGoGu}, C$_5$N$^-$ \cite{08CeGuAg}, and CN$^-$ \cite{10AgCeGu.cnm}. 
The majority of these species were first detected in a well observed Circumstellar Envelope
(CSE) IRC+10216, although these and other hydrocarbon anions were also discovered later on in other molecular clouds \cite{10SaShHi}.
As it is to be expected, the study and search for interstellar anions of both simple and increasingly more complex structural properties is
still current and relevant. More specifically, a simple species like C$_2$H$^-$ ($X^1\Sigma^-$), has been expected to be amenable to
observation with the new astrophysical instruments such as ALMA, especially since it had been already observed as a stable molecule in
laboratory experiments \cite{07BrGoGu,08Amxxxx}. Furthermore, 
its parent neutral form C$_2$H ($X^2\Sigma^+$) has already been a well-known astrophysical
molecule discovered by observation as early as 1974 \cite{74TuKuTh}, thereby suggesting that the corresponding C$_2$H$^-$ anion should also be
present, although perhaps as only a very low-abundance species, a fact justified in terms of its high chemical reactivity and therefore
expected rapid destruction upon formation. Actual current numerical models predict, in fact, that the formation of larger carbon chain anions
 should be in any case more probable than that for similar, but smaller chains as the C$_2$H$^-$ \cite{01BaSnBi}, 
hence somehow supporting the difficulties for its observation. 

The detection of carbon chain anions and parent neutral species in various astronomical environments has stimulated many computational  studies of their
collisions with He and H$_2$ to obtain rotationally inelastic excitation and de-excitation rates. Examples include the series
C$_2$H/C$_2$H$^-$ \cite{12SpFeNa,18Daxxxx,17DuLiSp,12DuSpSe} , C$_4$H$^-$ \cite{19SeDaDu}, C$_6$H/C$_6$H$^-$ \cite{17WaLiDu,16WaLiDu}
and CN/CN$^{-}$ \cite{10LiSpFe.cnm,11LiKlxx.cnm,12KaLiKl.cnm,13KaLiKl.cnm,11KlLixx.cnm,20GoMaWe},
 C$_3$N/C$_3$N$^-$ \cite{17LaStHa,19LaStHa,18TcMoNs}.

In spite of the fact that the actual origin of many of the hydrocarbon anions has not yet been solved, it is generally accepted that gas-phase
processes are crucial for their formation and therefore the many observed hydrocarbon radicals C$_n$H may also be the main precursors of the
formation of C$_n$H$^-$ anions through electron attachment or association processes, whereas associative detachment processes would contribute
to the generation of initial, neutral C$_n$H species. As an example, a new mechanism for the formation of C$_2$H$^-$ from C$_2$H$_2$ has been
proposed not long ago \cite{08CoMiWa} from laboratory experiments. Recent experiments in our group have also investigated reactions of
C$_2$H$^-$ with C$_2$H$_2$ as a means of forming larger C$_{2n}$H$^-$ and C$_{2n}$ chains \cite{19BaTiMe}.

Along similar lines, another small C-bearing molecule, the neutral C$_2$N radical ($X^2\Pi_r$) has been also detected earlier on in the
interstellar medium \cite{14AnZixx}, where the molecule was observed at the 1 -- 2 mK level toward the same 
circumstellar envelope (CSE) of IRC + 10216 already mentioned earlier, using the facilities of the Arizona Radio Observatory (ARO). 
Lambda doublets of the $J = 4.5 \rightarrow 3.5$ and
the $J = 6.5 \rightarrow 5.5$ transitions at 106 GHz and 154 GHz in the $\Omega = 1/2$ ladder were measured with the ARO 12 m telescope, as
well as the $J = 9.5 \rightarrow 8.5$ lines near 225 GHz, using the ARO Sub-Millimeter Telescope (SMT) \cite{14AnZixx}. 
Considering other species in the same environment, it is interesting to note that the [CN]/[C$_2$N]/[C$_3$N] abundance ratio was 
found to be $\sim 500/1/50$, thus indicating again one could expect a rather low abundance for the neutral radical C$_2$N. For the 
corresponding anionic counterpart, however, no observational evidence within the same CSE has been reported thus far, also suggesting a 
low abundance of the latter once formed within that chemical network.

In spite of this absence of direct detection, both the above anionic molecules, C$_2$N$^-$ and C$_2$H$^-$, have been the object of several
laboratory studies which have analysed in some detail the photo-detachment (PD) mechanisms of these species and several other structural
properties \cite{84KaSuSa,09GeYaNe,09HuLexx,91ErLixx.c2m,07ZhGeNe,13SeHoxx}.
Furthermore, in general terms we should note here that possible astrophysical abundances of either observed or not yet
observed species have to be understood in terms of molecular stabilities, reaction probabilities and of both radiative and collisional
excitations and relaxation of internal molecular modes: the accurate knowledge of all these facts can indeed help us to better explain the
existence of a molecule and the probability of it being observed. The molecular stability and the spectroscopic properties of both the 
C$_2$H$^-$ and C$_2$N$^-$ anions have been studied in various earlier investigations \cite{84KaSuSa,09GeYaNe,13SeHoxx}, 
while the modelling of molecular emission in the
ISM environments where they could be expected to exist requires collisional rate coefficients with the most abundant interstellar species like
He and H$_2$. Collisional data have been already presented for the C$_2$H$^-$ anion interacting with He \cite{12DuSpSe}, while no
corresponding calculations, as far as we know, exist for the collisional rotation-state changes of C$_2$N$^-$ interacting with He. 
The actual \textit{ab initio} PES is also not known for that system. The quantum dynamics of both these anions, besides not being 
extensively studied under CSE conditions, has also been only partially discussed under the operating low temperatures of ion cold traps
\cite{19GiGoMa.c2hm}. The present work is therefore directed to acquiring novel knowledge about the quantum dynamics of C$_2$N$^-$ in
collision with He under both astrochemical and cold ion trap conditions, further implementing a comparison between its dynamical 
behaviour and that of the C$_2$H$^-$ polar anion under similar conditions.

The following section will only briefly remind readers about the features of the anisotropic potential energy surface involving C$_2$H$^-$ 
and He atoms, since it has been presented and analysed already in earlier work \cite{12DuSpSe}. It will instead present in greater detail 
the new calculations of the \textit{ab initio} points related to the Rigid Rotor (RR) PES associated to the interaction between C$_2$N$^-$
($X^3\Sigma^-$) and the neutral He atom. The two interaction potentials will then be compared and their level of spatial anisotropy will be
analysed and discussed.

Section \ref{sec:rates} will present the state-changing rotationally inelastic cross sections for both systems and discuss the effects from 
spin-spin and spin-rotation structural features on the C$_2$N$^-$ system \textit{via \`a vis} the C$_2$H$^-$ ($X^1\Sigma^+$) system. 
The corresponding inelastic rates at the temperatures of interest will also be presented, compared and discussed. 
Section \ref{sec:traps} shall examine the possible evolutionary dynamics of C$_2$N$^-$ anions under the conditions of a cold trap when they
are undergoing laser-driven photo-detachment processes. A comparison with the corresponding behaviour of the C$_2$H$^-$ system will also be 
presented and discussed. Conclusions will be given in section \ref{sec:conclusions}.

\section{\label{sec:pes}Features of the \textit{ab initio} interactions of C$_2$H$^-$ and C$_2$N$^-$ with He atoms}

The electronic ground state of C$_2$N$^-$ was taken to be ($X^3\Sigma^-$) as indicated  in earlier
work \cite{09GeYaNe}. The molecular geometry of this linear anion is $r_{CC} = 1.344$ \AA\ and $r_{CN} = 1.207$ \AA\ and was given 
in the experiments from
Garand \textit{et al.} \cite{09GeYaNe}. We have optimized the molecular geometry using the MRCI + Davidson 
correction \cite{MOLPRO,MOLPRO_brief} employing the aug-cc-pVQZ \cite{94WoDuxx}
basis set obtaining $r_{CC} =  1.360$ \AA\ and $r_{CN} = 1.212$ \AA\ . The \textit{ab initio} calculations of the 2D grid of points in 
Jacobi coordinates (R, $\theta$) were carried out with the program package MOLPRO 2012 \cite{MOLPRO,MOLPRO_brief}. For
all the present calculations we have employed the internally-contracted multi-reference configuration interaction method (IC-MRCI) 
\cite{88WeKnxx,11ShKnWe.LM}, using the aug-cc-pVQZ basis set \cite{94WoDuxx} on all atoms. 
The reference space for the MRCI calculations consists of a complete active space by
distributing 14 electrons in 15 orbitals. All single and double excitations from the reference configurations are included in the variational
calculation. The effect of quadruple excitations is estimated via the Davidson correction \cite{MOLPRO,MOLPRO_brief}. Since the MRCI approach does not provide a size-consistent method, we have included a correction for size-consistency by using the Davidson correction that estimates the contributions of quadrupole excitations as discussed in references \cite{74Langdav, 82Buenpeye}. The Basis-Set-Superposition-Error (BSSE) correction was not included in this study because this procedure is not well defined when multi-reference methods are employed. In any event, our previous experience with small anions interacting with He has shown us that the corresponding rigid-rotor PES is only marginally modified by BSSE corrections in the well regions and on the onset of the repulsive walls. We therefore expect that it would also be not very significant in the present case.
We further checked the quality of the asymptotic form of the interaction by calculating the corresponding potential term for the situation where He approaches with an angle of 90\textdegree  to the molecular axis. In this situation the terms of the potential involving the higher order Legendre polynomial should be zero (by symmetry) and the spherical part should dominate. The computed value of the coefficient for the asymptotic potential gives us the value of the spherical polarizability of the He atom, which comes out to be 1.379  $a_0^{3}$. This value is in good agreement with a recent result from QED calculations\cite{20puchszale}  of 1.38376078   .

To further test possible differences coming from our different post-Hartree-Fock methods we have used, we have carried out a a comparison between MRCI(Q) and CCSD-T calculations for the approach of the He partner  at an angle of 90 degrees with respect to the axis of the
C$_2$N$^-$ ion. In the comparison the  potential energy curves for CCSD-T and MRCI(Q)
are set to zero at 25 Angstrom. We found that both curves have their minimum at 3.5 Angstrom, with  the largest difference between the two curves occurring  at the location of the minimum configuration: -59.9 cm$^{-1}$  for CCSD-T compared to -54.5 cm$^{-1}$ for MRCI(Q).

\begin{figure}
\includegraphics[width=0.4\textwidth]{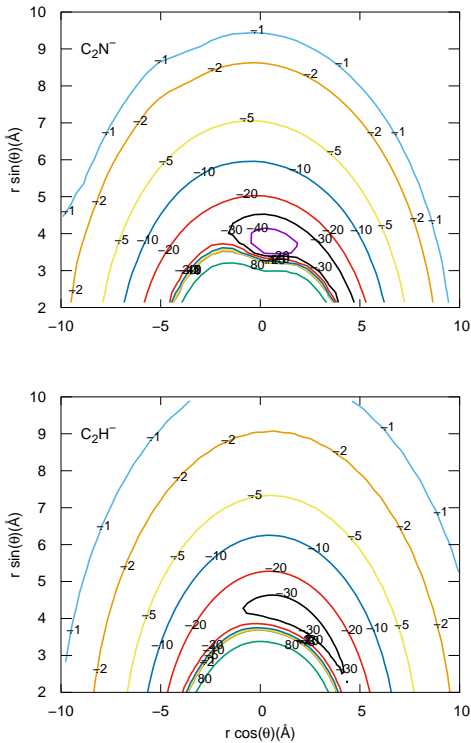}% Here is how to import EPS art
\caption{\label{fig:pes} Computed PESs for the two molecular anions of the present study. The data are presented as in-plane maps in 2D with
$r \cos{\theta} $ and $r \sin{\theta} $ as coordinates. Energy levels in cm$^{-1}$. Upper panel: C$_2$N$^-$ from present calculations; 
lower panel: C$_2$H$^-$ from ref. \cite{12DuSpSe}.}
\end{figure}

The angular grid involved calculations of
radial `cuts' every 5\textdegree, from 0\textdegree to 180\textdegree. The radial grid included a higher density of points around the
various minima regions at each selected angle. A total of 44 radial points were used between 2.3 and 26 {\AA}. We have not included the basis-set-superposition-error procedure because such a procedure turns out  not to be well defined for multi-reference methods.
The global minimum for the complex was found at a distance of around 3.545  {\AA}    from the centre-of-mass, located at an angle of around
 80\textdegree, with a well depth of around 
58 cm$^{-1}$. The data shown by the two panels of Figure~\ref{fig:pes} provide a pictorial view of the new 
PES calculated for the C$_2$N$^-$/He 
system (upper panel), while it also shows for comparison the PES already calculated in earlier work for the similar anion of C$_2$H$^-$
interacting with He \cite{12DuSpSe}.  One  sees in the two panels similarities between the two interaction potentials since both exhibit 
the presence of a minimum energy region of their complex with He. The more shallow well of the lower panel is extensively off the 
C$_{\rm 2v}$ geometry, while the deeper well on the upper panel appears to be more localized and closer to the center-of-mass than it 
is the case in the lower panel. Given the larger number of electrons in the C$_2$N$^-$ case, the corresponding well is therefore deeper 
than that for the C$_2$H$^-$/He complex \cite{12DuSpSe} and more localized in space. Both potentials will be asymptotically driven by the 
polarizability term involving the spherical, dipole polarizability of the He partner and therefore will behave very similarly in their 
long-range regions because of it. The differences in their short-range features, however, will be reflected later in the size 
differences between their collisional state-changing probabilities discussed in the next Section.

\begin{figure}
\includegraphics[width=0.4\textwidth]{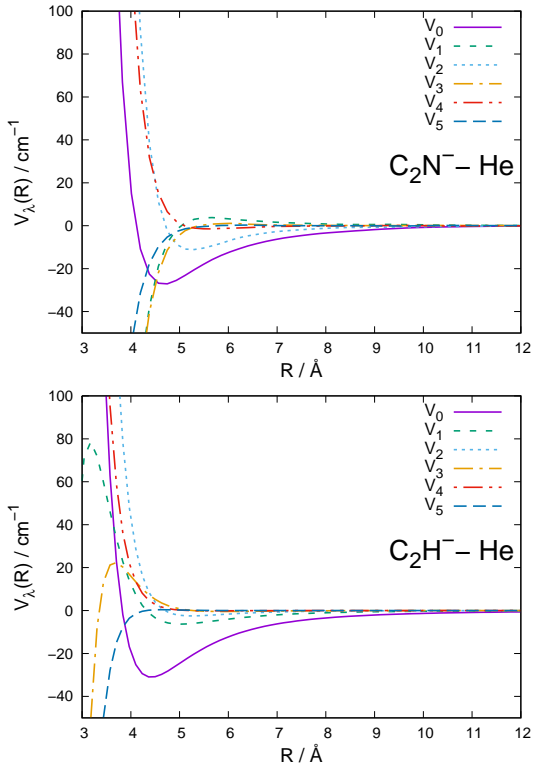}% Here is how to import EPS art
\caption{\label{fig:lambda} Computed multipolar coefficients calculated from the initial PES data. Upper panel: lower values of the 
dominant radial coefficients for the C$_2$N$^-$ anion; lower panel: same data but for the C$_2$H$^-$ anion. 
Energy values in cm$^{-1}$ and distances in \AA\ .}
\end{figure}

The localization of the excess charge provided by the extra bound electron of the anion is also an interesting item for the new 
C$_2$N$^-$/He PES presented here. 
The Mulliken charges in the asymptotic situation (e.g. with the helium at a distance of 26.55  {\AA}     from the
center-of-mass of the anionic target) turn out to be as follows (when computed with the MRCI / aug-cc-pVQZ of the post-HF treatment): 
C$_1$ = -0.59049; C$_2$ = +0.22532 ; N$_1$ = -0.63483. Here C$_2$ is the central carbon atom, so that the geometry of
C$_2$N$^-$ is given as: C$_1$-C$_2$-N$_1$ , with no extra charge on the interacting He atom, as expected. 
The direction and value of the dipole
moment are therefore given as: 2.1818 Debye, placed along the positive direction of the molecular z-axis from the C$_1$-end of the molecule, 
a result from the delicate balance between excess charges in this molecule.One should also note that the approach of the He atom along the direction of 0\textdegree  is on the N-side of the 
C$_2$N$^-$ anion. The same angle corresponds to the approach on the H-side of the C$_2$H$^-$.
Another type of presentation of the PESs for both systems, to be used in the scattering calculations below, is obtained by numerically
 generating the radial coefficients of the multipolar expansion of the Rigid Rotor (RR) 2D potential energy surfaces:
\begin{equation}
V(r=r_e , R, \theta) = \sum_{\lambda} V_{\lambda}(R)P_{\lambda}(\cos\theta)
\end{equation}
where $r_e$ is the geometry of the equilibrium structure of the anion and the sum over the contributing 
$\lambda$ values went up to 19, although only the dominant, stronger terms are shown in Figure~\ref{fig:lambda}. The panels of that figure
also compare the present findings for C$_2$N$^-$ with the earlier data for C$_2$H$^-$ \cite{12DuSpSe}. The expansion coefficients for the 
C$_2$N$^-$-He PES are provided in the supplementary information.

\begin{figure}
\includegraphics[angle=270,width=0.6\textwidth]{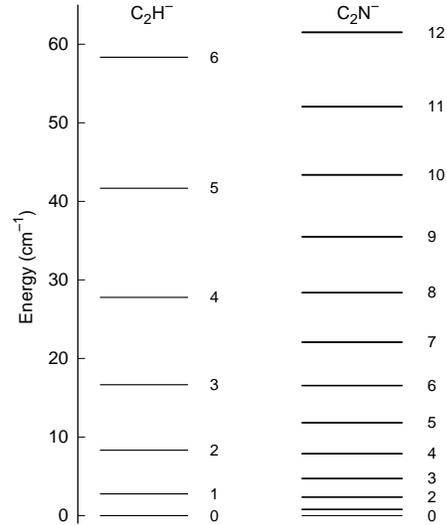}% Here is how to import EPS art
\caption{\label{fig:rot} Schematic location of the rotational energy levels for the molecular anion of the present study. 
For the case of C$_2$N$^-$ ($X^3\Sigma^-$) only the pseudo-singlet levels without splitting effects are shown. 
See main text for further details.}
\end{figure}

The radial coefficients presented in Figure \ref{fig:lambda} underline once more both the general similarities between the two systems and
important differences which will vary their dynamical behaviour we shall analyse below. If we look at the spatial anisotropy of the
 multipolar coefficients we see that the spherical terms 
($\lambda$ = 0) are very similar to each other in both strength and spatial extension around their well regions. On the other hand, the 
radial coefficients with $\lambda$ = 1, 2 and 4 are all uniformly repulsive in the short-range region 
for C$_2$H$^-$, while only those for $\lambda$= 2 and 4 are so for the
C$_2$N$^-$ anion. These differences indicate that the $\Delta N$ transitions between rotational levels, those for which the acting torques
from the anisotropic PES coefficients will be stronger, could be those with $\Delta N$  with odd values for C$_2$N$^-$  and with even
values for the  C$_2$H$^-$. The presence of such possible propensity effects will be further discussed when analysing the computed inelastic
cross sections and rates in the next section. Another important element of distinction/similarity 
are the relative spacings in energy between the rotational states for both 
systems which are presented in Figure~\ref{fig:rot}.

To simplify the comparison we report only the energy spacing between $N$-level separations, 
without showing the spin-spin and spin-rotation splitting for 
C$_2$N$^-$. The sizes of those splitting constants are not known or available as yet and we will be discussing further in the next 
Section how their effects will be included in the dynamics of the present study. Their energy splittings would in any event not be 
visible on the chosen energy scale in that Figure. We clearly see, however, that the density of states over the examined range of about 60 
cm$^{-1}$, which should cover most of the relevant energy range of  the CSE environments and of the cold ion traps operation, is 
dramatically different between the two systems, with the C$_2$N$^-$ anion showing twice as many states being accessible within that 
energy range. The markedly higher `crowding'  of rotational states per unit energy for the latter molecule translates into a higher
number of them being significantly populated at equilibrium temperatures in either environments. 

\begin{figure}
\includegraphics[trim=0 8cm 0 0,scale=0.3,angle=270]{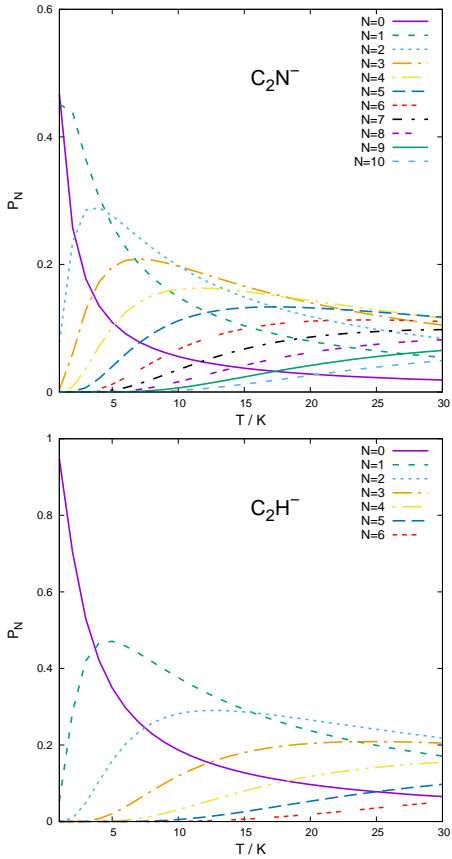}% Here is how to import EPS art
\caption{\label{fig:boltz} Steady-state distributions of relative molecular populations of rotational levels for temperatures up to 30 K. 
Left panel: the C$_2$N$^-$ anion; right panel: the C$_2$H$^-$ anion. See main text for further discussion.}
\end{figure}

The use of a Local Thermal Equilibrium 
(LTE) population scheme is often employed as a starting condition of ISM kinetic modellings. Additionally, photo-detachment experiments in 
cold traps usually start, as we shall further discuss in a later Section, with thermal equilibrium of the molecular internal states 
being achieved at the temperature of the trap.
A comparison between steady-state rotational level populations, over the range of temperatures of interest, is reported for the two anions 
 by the two panels in Figure~\ref{fig:boltz}.  We clearly see  in them that, as an example, at
temperatures in a cold trap around 25 K the C$_2$N$^-$ anion in the left panel significantly populates levels up to $N=10$, 
while the C$_2$H$^-$ anion on the right panel only has states up to $N=6$ significantly populated. Such differences of behaviour will be
further discussed in the next two Sections when the inelastic cross sections and rates will be computed first  to numerical convergence and
then the possible dynamics of laser-induced PD processes in traps will be analysed at  low temperatures.

Since the efficiency of state-changing collisional processes for rotational states is affected, among other features also discussed below, 
by the size of the energy gaps between the involved levels, we shall therefore expect that the marked differences shown by 
Figure ~\ref{fig:rot} will be reflected in differences between inelastic rates  for the two anions.

\section{\label{sec:rates} Quantum calculations of state-changing rotationally inelastic cross sections and rates}

The inelastic cross sections involving collisional transitions in both the title molecules were calculated using our in-house multichannel
quantum scattering code ASPIN, which we have already described in many earlier publications of our 
group \cite{08LoBoGi,03MaBoGi,17HeGiWe,06GoBoGi}
and therefore will not be repeated here in detail. Interested readers are referred to the above publications for consultation. 

The ground state of the C$_2$N$^-$ anion is $^3 \Sigma^- $, the same as the OH$^+$ molecule for which we have previously studied 
low-temperature collisions with He in detail \cite{07GoBoGi,18GoWeGi}. 
The conclusions of this previous work inform the calculations presented here.
In the OH$^+$ ($^3 \Sigma^- $) state, we have three levels for each total angular momentum $\ge 1$: the rotational levels are in fact 
split by spin-spin and spin-rotation coupling effects. In the pure Hund's case (b) the electronic spin momentum $S$ couples with the nuclear
rotational angular momentum $N$  to form the total angular momentum $j$, given 
by $j = N + S$   \cite{07GoBoGi,18GoWeGi}.
As a consequence of that coupling term, the rotational levels of the molecule may be labelled not only by the above quantum numbers defined
in the Hund's case (b) but also by their parity index $\varepsilon$. The levels in molecules of odd multiplicity with parity index equal 
to +1 are labelled {\bf e}, and those with parity index equal to -1 are labelled {\bf f} \cite{07GoBoGi}. However, we shall omit this index in 
the following discussion since we shall not be using it in our analysis.

The extensive work we have done on collision of He with the OH$^+$ ($^3 \Sigma^- $) molecule \cite{07GoBoGi,18GoWeGi} 
have already shown that the $\Delta N \ne 0$
transitions involve much larger energy values than those which cause changes in the spin quantum number $S$ that is responsible for the
separations between $j$ states. Thus, one can say that rotational quenching transitions have much larger energy gaps than those which simply
cause spin-flipping processes within each $N$-labelled manifold.
In a full close-coupling (CC) approach to the quantum dynamics the molecular Hamiltonian includes, in addition to the rotational 
contribution, a nuclear coupling contribution which induces hyperfine energy splitting. These splittings are however lower than typically
10$^{-3}$ cm$^{-1}$, i.e. they are much lower than the rotational spacings investigated in this work, as we shall
further show later in this paper. In such situation, a possible approach   
would  be to neglect the hyperfine splitting and to decouple the spin wave functions from the rotational wave functions, 
using an angular momenta recoupling scheme (e.g. see: Alexander \& Dagdigian \cite{83AlDaxx})  which simplifies considerably
the dynamics of the problem. The latter is then reduced to solving the simpler spin-less CC equations associated to the more 
usual ${}^1 \Sigma$ case. This approximate approach of treating the target as a pseudo-singlet case has been found in earlier works of our
group \cite{17HeGiWe,18GoWeGi} to yield results very close to the exact cross sections. Hence the  `recoupling approach' outlined 
above will be considered as our reference approach in the following calculations, where
our results will be compared with those carried out for the exact ${}^1 \Sigma$ case of the C$_2$H$^-$ molecular anion.

For C$_2$N$^-$-He (reduced mass $\mu = 3.619$ amu), scattering calculations were carried out for collision energies between 
1 and 500 cm$^{-1}$ using steps of: 0.05 cm$^{-1}$ for 0 to 6
cm$^{-1}$, 0.1$^{-1}$, for 6-100 cm$^{-1}$, 0.2 cm$^{-1}$ for 100-200 cm$^{-1}$, 1.0 cm$^{-1}$ for 200-400 cm$^{-1}$ and of 2 cm$^{-1}$ for
400-500 cm$^{-1}$. 
This fine energy grid was used to ensure that important features such as the many resonances appearing in the cross sections were 
accurately accounted for and their contributions correctly included when the corresponding rates were calculated, as discussed below. 
The CC equations were propagated between 1.7 and 
100.0 {\AA} in 2000 steps using the log-derivative propagator \cite{86Maxxxx.c2m} up to 60 {\AA} and the 
variable-phase method at larger distances \cite{03MaBoGi} up to 100 {\AA}.
The potential energy was interpolated between calculated
$V_{\lambda}(r_{\rm{eq}}|R)$ values using a cubic spline and extrapolated below and above the \textit{ab initio} grid
using linear and polynomial functions respectively as implemented in ASPIN \cite{08LoBoGi}. 
As our \textit{ab initio} grid goes to 26 {\AA} and the scattering energies of 
interest are not in the ultracold or high energy regimes, the details of the potential extrapolation have a negligible effect on the
scattering cross sections. To test this point, however, we have always extended the radial integration with an extrapolated potential of the R$^{-4}$- type out to 100 {\AA} or more to reach convergence of the final cross sections. We have also found that to employ the correct dipole polarizability coefficient for the extrapolated potential instead of the value produced by ASPIN from the ab initio potential  only changed the final cross sections by less than 1.0{\%}, and only at the lowest energies.
The rotational basis set was increased with increasing energy from $N=20$ to $N=36$ at the highest energy considered.
The number of partial waves was also increased with increasing energy reaching $J = 96$ at the highest energy. 
Inelastic cross sections were computed for all transitions between $N = 0$ to $N = 15$, which was deemed to be sufficient to model 
buffer gas dynamics in a cold trap up to about 50 K (see below), while the same range of levels is expected to be the one most 
significantly populated during low-energy collisional exchanges with He atoms within ISM environments dynamical conditions.
Scattering calculations for C$_2$H$^-$ were carried out by us in a former study \cite{19GiGoMa.c2hm} using the PES of 
Dumouchel \textit{et al.} as discussed above \cite{12DuSpSe}.

Another quantity that we shall compute from the cross sections is the state-to-state inelastic rotational rates over a range of 
temperatures from thresholds up to about 100 K, to cover the range of $T$ values expected to be significant for processes in the ISM
environments we are discussing here.
Hence, once the state-to-state inelastic integral cross sections are known, the rotationally inelastic rate constants 
$k_{N \rightarrow N'} (T)$ can be evaluated as the convolution of the cross sections over a Boltzmann distribution of the 
relative collision energies. In the equation below, all quantities are given in atomic units:
\begin{equation}
k_{N \rightarrow N'} (T) = 
\left(  \frac{8}{\pi \mu k_b^3 T^3}   \right)^{1/2}
\int_0^{\infty} E \sigma_{N \rightarrow N'}(E) e^{-E/k_B T} dE  
\end{equation}
Both the calculations of inelastic cross sections and of their corresponding rates over the range of temperatures mentioned earlier 
have been carried out using the new interaction PES for the C$_2$N$^-$/He system and via the earlier PES already computed for 
the C$_2$H$^-$/He system \cite{12DuSpSe}.  
The comparison between our findings for these two systems will be  carried out below
by presenting our computed data.

\begin{figure}
\includegraphics[angle=270,width=0.47\textwidth]{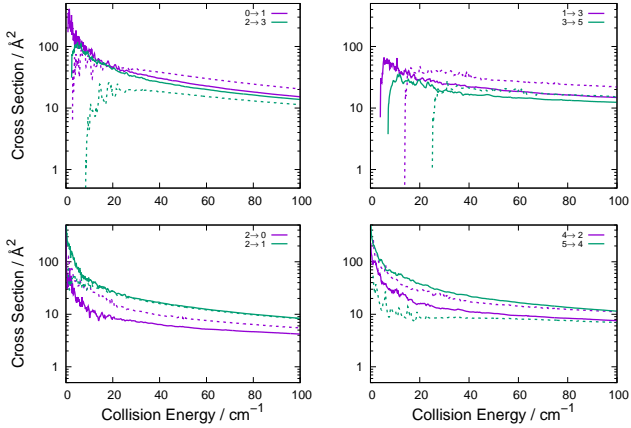}% Here is how to import EPS art
\caption{\label{fig:cross} Rotationally inelastic cross sections for the C$_2$N$^-$ (solid lines) and the C$_2$H$^-$ (dashed lines) 
anions. The upper panels report excitation processes while the lower panels describe de-excitation processes. See main text for further
discussion.}
\end{figure}

The present results for the inelastic cross sections, over a range of 100 cm$^{-1}$, are reported 
in the four panels of Figure ~\ref{fig:cross}. The upper two panels show excitation processes while the lower two panels present 
de-excitation cross sections for both systems. The solid lines refer to the calculations for C$_2$N$^-$ while the dashed lines are
those for C$_2$H$^-$.

It is interesting to notice that in the low-energy range up to about 20 cm$^{-1}$ all of the cross sections reported in the four panels 
show evidence of resonances effects in both systems. The relative  strength of the interaction potentials and the presence of many 
more rotational states available for the C$_2$N$^-$ makes the cross sections for the latter anion exhibit a much denser sequence of 
resonant features, as expected from the larger mass of the molecular partner (hence a larger number of partial waves contributing to 
shape resonances) and form the stronger coupling with the nearby closed channels during the scattering (hence the presence of Feshbach
resonances). Since no experiments are as yet available on the scattering behaviour of these anions, we shall not carry out any detailed
analysis of the features seen in our calculations. They will however contribute to changing the size of the related inelastic rates
we shall further discuss later, and will be taken into account in our calculations.

In the upper two panels where excitation processes are reported, we see that in the left panel,
showing  $\Delta N$=1 transitions, at energies beyond the resonance regions, the excitation from the 
lowest rotational state of C$_2$H$^-$ are the largest 
over the whole range, in spite of the fact that the amount of energy transferred between levels is also the largest. 
Two structural factors could contribute to this dynamical difference: \textit{(i)} the smaller value by about 30\% of the reduced mass of the 
C$_2$H$^-$ which affects relative sizes of wavevectors, and \textit{(ii)} the presence for only this molecule of an attractive well in the 
radial coefficient with $\lambda$= 1 which directly couples the levels involved in that transition during collision. The marked 
increase in energy gap when one considers transitions still with $\Delta N$=1 but between excited states (i.e. between the $j = 2$ 
and $j = 3$ levels) for the C$_2$H$^-$ system in the same upper-left panel, is the dominant factor which is causes 
that inelastic cross section to be the smallest. On the other hand, the two excitation cross sections involving the same levels but 
for the C$_2$N$^-$ anion which are in that same panel are closer to each other. This is due to the much smaller energy gap between 
its levels which does not offset the greater strength of the interaction potential for the latter molecule compared to the former, hence 
making the C$_2$N$^-$ anion more efficiently excited by collisions with He to its upper rotational states.

The panel in the upper-right part of Figure \ref{fig:cross} shows now excitation processes involving $\Delta N$=2 transitions. 
In the low-energy resonance regions the cross sections are uniformly smaller than those discussed before for the C$_2$N$^-$ anion, 
while those for the  C$_2$H$^-$ anion are instead larger in size and also larger than those pertaining to the other anion. 
This difference can be linked to the differences in behaviour between their corresponding multipolar coefficients, 
as shown in Figure ~\ref{fig:lambda}. The C$_2$H$^-$ partner presents both terms with $\lambda$= 1 and $\lambda$= 2 with marked attractive
wells and steeply repulsive walls, while the other anion only shows that behaviour for the  $\lambda$= 2 coupling term. It therefore follows
that both direct and indirect potential coupling effects are dynamically efficient for the C$_2$H$^-$ anion, while only one is effective 
for C$_2$N$^-$.

When turning to the collisional de-excitation processes in the lower panels of Figure \ref{fig:cross}, we see in the left panel that the 
transitions with $\Delta N=-1$ for the C$_2$N$^-$ and the C$_2$H$^-$ anions yield markedly larger cross sections than those for 
de-excitation transitions with $\Delta N=-2$, a behaviour attributed again to the role played by the increase of the energy gaps in the 
case of the latter processes. Also the differences in the coupling strength between the two PESs discussed earlier cause the 
$\Delta N=-2$ transitions for C$_2$N$^-$ to be the smallest of them all. It is also interesting to note that, in the lower-left panel, 
the smallest of the cross sections is that for the $\Delta N=-1$ transition of the C$_2$H$^-$ partner. The much larger energy gain 
transferred into relative energy between collision partners makes this system have a reduced interaction time that in turn makes 
sudden collisions more effective. Since such gain is larger for  $\Delta N=-2$
processes than for the $\Delta N=-1$ transition, it makes sense that the cross sections for the latter are smaller than those for the 
former process when C$_2$H$^-$ is concerned. The differences in energy gains are however much smaller for C$_2$N$^-$, where the 
differences between multipolar coefficients play a more important role than the smaller changes in the interaction times, thus
 making the $\Delta N=-1$ transitions to be the larger ones.

On the whole, the present quantum calculations of the relative state-changing collision probabilities for the two title systems 
indicate a clear similarity of general behaviour and of relative efficiency for such processes at the energies of interest for ISM
conditions, although changes linked to their structural features can distinguish between their interactions with the He atom and  play a
significant role in their relative dynamics.

\begin{figure}
\includegraphics[angle=270,width=0.47\textwidth]{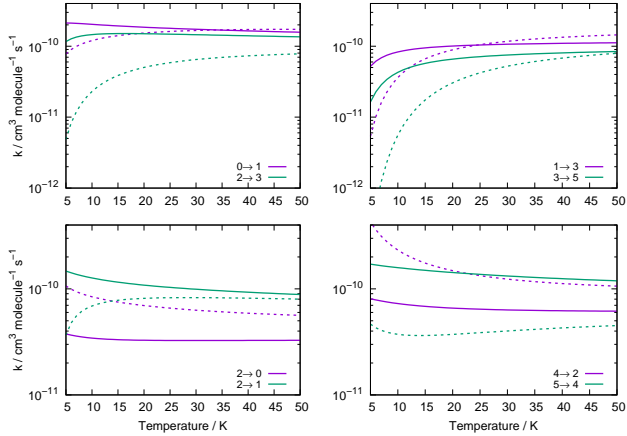}% Here is how to import EPS art
\caption{\label{fig:rates} Rotationally inelastic collisional rates for the C$_2$N$^-$( solid lines) and the C$_2$H$^-$ (dashed lines) 
anions. The upper panels report excitation processes while the lower panels describe de-excitation processes. The range of temperatures 
goes up to 50 K. See main text for further discussion.}
\end{figure}

The results reported by Figure ~\ref{fig:rates} indicate in the behaviour of the rotationally
inelastic collisional rates over a range of temperature covering values of interest for cold ion traps and also for ISM environments 
where the present anions are expected to be present. As with the cross sections of Figure \ref{fig:cross}, the upper two panels 
report excitation processes while the lower two panels show the behaviour of the de-excitation processes. Solid lines refer to the 
C$_2$N$^-$ anion while the dashed lines are for the C$_2$H$^-$  case.

It is interesting to note, when looking at the results in the two upper panels, that the differences in size between rates for 
$\Delta N$=1 and those for $\Delta N$=2 are fairly small in the case of the C$_2$N$^-$ target, while are markedly larger for C$_2$H$^-$ 
 as a target. This is again due to the greater effects of the energy gaps between rotational states which exist between the 
two anions, as shown by Figure ~\ref{fig:rot}. These types of differences are also present in the de-excitation rates 
reported by the two lower panels , where again we see that the differences in size of the rates for each of the anion are all 
within less than one order of magnitude. As expected, we see that the rates follow closely the indicated differences discussed 
earlier between cross sections for both systems.

\begin{figure}
\includegraphics[angle=270,width=0.47\textwidth]{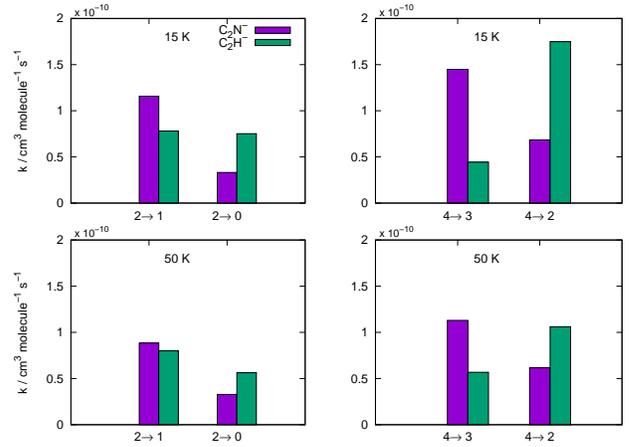}% Here is how to import EPS art
\caption{\label{fig:ratesticks} Rotationally inelastic collisional quenching rates for the C$_2$N$^-$ and the C$_2$H$^-$ 
 anions for $N=2$ (left panel) and $N=4$ (right panel) at 15 K (top panels) and 50 K (bottom panels).
 See main text for further discussion.}
\end{figure}

The  data reported by the four panels of Figure ~\ref{fig:ratesticks} make a pictorial  comparison between sizes of specific  
rate values  at two different temperatures: at 15 K in the two upper panels and at 50 K in the two lower panels. 
They confirm what was discussed earlier and underline the delicate interplay between the structural and dynamical differences controlling  the 
state-changing collisions between rotational states in the two anions.

The upper two panels of Figure \ref{fig:ratesticks} compare two  de-excitation rates 
for two different initial states: $N$=2 in the left panel and $N$=4 in the right panel. The following comments can be made : 

\textit{(i)} the rates for the C$_2$N$^-$ anion are larger for  $\Delta N=-1$ processes, while becoming smaller than 
those for C$_2$H$^-$ for $\Delta N=-2$ processes. The switching is related to the differences in 
multipolar coefficients discussed  earlier.

\textit{(ii)} for the same types of transitions at the higher temperature shown in the lower two panels, we see that the 
inversion of relative magnitudes when going from $\Delta N=-1$ processes to the $\Delta N=-2$ is still present and linked 
to different coupling strengths of their  state-changing dynamics.

\begin{figure}
\includegraphics[angle=270,width=0.47\textwidth]{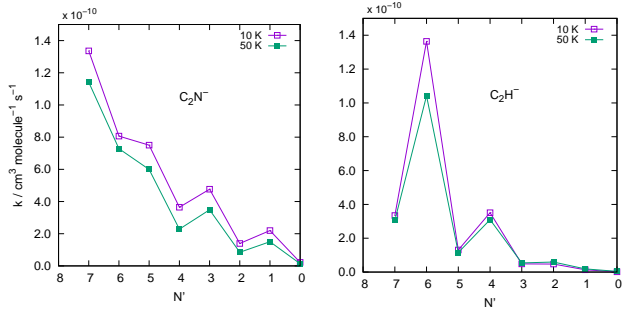}% Here is how to import EPS art
\caption{\label{fig:j8rates} Size changes of rotationally inelastic collisional rates for the C$_2$N$^-$ (left panel) and the C$_2$H$^-$
 (right panel)  anions. The two curves in each panel refer to two different temperatures as reported in the captions. The rates 
are calculated from the $N=8$ state in each molecule, at the two temperatures shown, and the downward rates are given as a function of 
the final state for each anion. See main text for further discussion.}
\end{figure}

Indicators on possible propensity rules for the size-dependence of the cooling (de-excitation) 
rates from an initial rotational state  and down to different final levels are reported in the two panels of Figure ~\ref{fig:j8rates}. 
The variations in size of the  rates from the initial $N=8$  to all final levels below are shown at two
different temperatures.

Figure \ref{fig:j8rates} verifies directly what we have already discussed when analysing the cross sections and 
the  rates, i.e. that in the collisional cooling (de-excitation) cascades within each system are linked to the differences in 
the structural features of the interactions between the anionic partners and the He atom.
It is immediately clear that temperature does not play a significant role and that structural 
features are largely driving the differences. In  the case of the C$_2$N$^-$ anion, increasing the $\Delta N$ values 
down the ladder  uniformly reduces the size of the inelastic 
rates as  the energy gaps uniformly increase. 
Such differences are smaller than those for C$_2$H$^-$ where larger energy spacings occur between levels.
Within this decrease in size, however, we still see that for C$_2$N$^-$ the transitions involving odd $\Delta N$ 
values are larger that those with even values, while the opposite is true for the C$_2$H$^-$. 
This is again linked to the dynamical couplings induced by the attractive 
$\lambda$= 1, 3 and 5 odd terms of the multipolar expansion of the C$_2$N$^-$, as shown by Figure~\ref{fig:lambda}.
On the other hand, for the C$_2$H$^-$ anion we see an even more marked effect linked to the 
relative strength of the multipolar coefficients (see again Figure~\ref{fig:lambda}), since the $\lambda$=2 
and 4 potential coupling terms have a shallow attractive well in the outer radial range and become strongly repulsive around the turning point 
values at the energies of interest. We therefore see that they  induce rotational torques during dynamics which would 
favour transitions into final states where even $\Delta N$ values occur.

It  follows from the above considerations that collisional repopulation of rotational levels will take different preferential 
paths in the two title systems, a result which would suggest the final presence in the relevant environments of different populations 
for their molecular rotational states.

Rates for all transitions between $N=0$ and $N=8$ for C$_2$H$^-$ and between $N=0$ and $N=15$ for C$_2$N$^-$ 
are provided in the supplementary information for temperatures between 0 and 100 K.

\section{\label{sec:traps} Modelling Quantum dynamics of laser photo-detachment processes in cold traps}

One of the important processes which can take place in cold ion traps \cite{19GiGoMa.c2hm,18GiLaHe} where 
molecular anions can be confined, involves studying the internal state evolution under laser-induced detachment of the extra anionic electron.
The study photo-detachment of molecular anions is obviously more  complicated than for atomic PD processes due to the increasing number of accessible pathways which become available in the case of molecular targets. Molecules exhibit a higher density of lower- lying electronic states while one must further consider the increase in the density of states due to the additional  vibrational motion, i.e. ro-vibrational and pre-dissociative states of the initial anion are actively present as the bound electron is moved into the continuum  by the laser beam. The analysis of the PD mechanism for molecular  negative ions also provides an important tool for the studies of reaction dynamics, where one wants to employ the initial anionic partner in a reaction to be specifically produced in a given ro-vibrational state of its ground electronic state as the neutral is generated in the PD reaction.

This process can therefore be outlined as having to  follow this sequence of events: \textit{(i)} collisional population re-distribution among the rotational 
levels of the anion loaded in the trap and interacting with He, as the prevalent buffer gas, at a given trap temperature;  
\textit{(ii)} switching on of the 
photo-detaching laser after the previous rotational population equilibration has been achieved. One could then decide to vary the 
laser's operational wavelength to selectively depopulate different rotational states of the anion present in the trap after the step
\textit{(i)}, and producing the ejected electron either at threshold or away from it, or simply follow the total anion losses 
in the trap during the photo-detachment (PD) step;  \textit{(iii)}  test the dynamical effects from changing other operating
conditions such as: buffer gas density, trap equilibration temperature, 
laser power and laser wavelength in order to find the best choices for the PD processes tailored to the specific systems. 

To follow the above steps in modelling the PD process,  the state-changing collisional rates for the molecular anion of interest need to 
be known at different operating temperatures. Furthermore, we need to know as best as possible the PD rates (obtained from the PD cross
sections as functions of the initial
rotational state, as we have previously discussed \cite{19GiGoMa.c2hm,18GiLaHe}). 
As a comment about the calculated cross sections and rates for state-to-state
inelastic processes, it suffices to say that the dominant inelastic processes we found for the present systems are those for which the 
state-changing values of $\Delta N \pm 1, 2$ and for which the rates at a temperature of, say 20 K, are of the order of 10$^{-10}$ 
in units of cm$^3$ s$^{-1}$. As a comparison, the corresponding spontaneous emission Einstein A coefficients for the C$_2$H$^-$ 
anion ( in units of s$^{-1}$)   vary to be from about 10$^{-3}$ to 10$^{-5}$ times smaller than the collisional rate constants for this system \cite{19GiGoMa.c2hm}. 
We shall therefore chiefly consider only collision-induced effects in the present analysis.

For modelling the rotational population evolution dynamics, the master equations need to be solved using the collisional thermal 
rates already computed above, at each chosen trap temperature and for the range of selected He density which are expected 
to be experimentally achieved:

\begin{equation}
\frac{dn_i(t)}{dt} =
\sum_{j \ne i} n_j(t) C_{ji}(T)
- n_i(t)  \left( \sum_{i \ne j} P_{ij}(T) + K_i^{PD} \right) \;\;,
\end{equation}

The quantities $P_{ij}(T )$ are the rates for the destruction of level $i$, while its formation rates are given by the $C_{ji}(T)$ terms.
During the collisional step, i.e. before the laser is switched on, the coefficients are given as: $P_{ij}(T )= \eta_{He} k_{i \rightarrow j}
(T) $ and  $C_{ji}(T ) = \eta_{He} k_{j \rightarrow i} (T)$. These relationships  describe the  `collision-driven'  time evolution
process of thermalisation of  the relative populations of the rotational levels of the anion which are reached at the selected buffer gas
temperature and for a given density $\eta_{He}$ in the trap. Once thermalisation is reached and the PD laser is turned on, 
$K_i^{PD}$ is the additional destruction rate of the selected level $i$ caused by the PD laser. The set of rates $K_i^{PD}$ is 
critical in the experiments in general and for the present numerical simulations because they drive the destruction of both the 
population of one specific rotational level $i$ and of all the molecular ions which have been populating that specific state during the
previous thermalisation step in the traps. As the trapped anion is still undergoing collisions with the buffer gas, there is now competition 
between the laser-induced de-population of a rotational level and its collisional re-population in the trap.

In the experiments of this study these rates depend on the laser photon flux and on the overlap between the laser beam and the ion cloud
within the trap. Since these parameters, as well as the absolute values of the state-to-state PD cross sections, are presently unknown 
for the title molecular anions, we shall introduce a scaling parameter which we have already discussed in our earlier 
work \cite{19GiGoMa.c2hm,18GiLaHe,20MaNoGo,20SiNoMi} according to which the relation between the required rate and the 
estimated cross section is given by: 

\begin{equation}
K_{N}^{PD} = \alpha(\nu) \sigma_{N}^{PD}(\nu)
\label{eq.alpha}
\end{equation}

The above scaling parameter $\alpha(\nu)$ accounts globally for modelling the relative role of different 
features of the photo-detachment experiment which are effectively the laser-driven features of the PD dynamics. They include quantities 
such as laser-flux, spatial overlap, etc. which compete with the pure collisional rates that repopulate the anion's rotational levels 
via its interaction with the buffer gas. The specific scaling of that parameter within the modelling allows us to simulate either 
a `collision-dominated' situation or a `PD-dominated' situation whenever the laser strength is markedly varied by taking 
the  $\alpha(\nu)$ to be either equal to 1.0 or one or more orders of magnitude smaller, as shown in our earlier work
\cite{19GiGoMa.c2hm,18GiLaHe,20MaNoGo,20SiNoMi} 
and as  discussed below. The values of $\sigma_{N}^{PD}(\nu)$ can be obtained through an analysis which we have
presented before \cite{19GiGoMa.c2hm,18GiLaHe,20MaNoGo,20SiNoMi}. Briefly, the PD cross section is given as:

\begin{equation}
\sigma_{N}^{PD}(\nu) 
\propto 
\sum_{N`= 0}^{N_{\rm max}} \left|  C_{N 010}^{N' 0}  \right|^2 
(E - E_{th})^p \Theta(E - E_{th}) \;\; .
\label{eq:PD}
\end{equation}

Here $\Theta(E - E_{th})$ is a step function so that only transitions with $E_{th} < E$ are considered, and the Clebsch-Gordan coefficient
enforces the selection rule $\Delta J'' = \pm 1$ for the specific PD process under consideration. Since the rotational levels' 
relative populations initially also sum to one, they have limited effect on the relative sizes of the PD cross sections and the 
PD curve identified by the above equation. Hence, the most important factors are the selected value for the $p$ exponential 
parameter, a factor linked to the dipole selection rules of the photon-induced electron detachment and to the dominant partialwave  ( angular momentum) of the ejected electron, as discussed in  \cite{19GiGoMa.c2hm,18GiLaHe,20MaNoGo,20SiNoMi}. The actual wavelength $\nu$ of the laser source with respect to the threshold of the specific electron-detachment process is also an important feature in the discussion of the PD processes, as we shall further show below.

\begin{table}[h!]
\caption{\label{tab:photo1} Relative values of $\sigma^{\text{PD}}_{N}(\nu)$ for C$_2$N$^-$ as a function of the rotational state N, given by the first column, and the different $p$ values reported by the next three columns. The  value of $\nu$ is fixed to be that of the 
electron affinity (EA) at 2.7489 eV.}
\begin{tabular}{cccc}
 $N$ & p=0.5 & p=1.5 & p=2.5  \\
\hline
     &     &      &          \\
0    & 0.00 & 0.00 & 0.00    \\
1    & 0.23 & 0.03 & 0.003    \\
2    & 0.35 & 0.06 & 0.01    \\
3    & 0.44 & 0.11 & 0.03    \\
4    & 0.51 & 0.16 & 0.05     \\
5    & 0.58 & 0.22 & 0.08     \\
6    & 0.64 & 0.28 & 0.12     \\
7    & 0.69 & 0.35 & 0.18     \\
\hline
\end{tabular}
\end{table} 

\begin{table}[h!]
\caption{\label{tab:photo2} Same data as in the previous table.Relative values of $\sigma^{\text{PD}}_{N}(\nu)$ as a function of $p$ for $\nu$ = EA 
+ 50 cm$^{-1}$.}
\begin{tabular}{cccc}
 $N$ & p=0.5 & p=1.5 & p=2.5  \\
\hline
     &        &       &        \\
0    &  1.000 & 1.000 & 0.934   \\
1    &  0.999 & 0.999 & 0.935   \\
2    &  0.999 & 0.999 & 0.936  \\
3    &  0.999 & 0.999 & 0.938  \\
4    &  0.999 & 0.999 & 0.940   \\
5    &  0.998 & 0.999 & 0.942   \\
6    &  0.997 & 0.999 & 0.946   \\
7    &  0.996 & 0.999 & 0.947    \\
\hline
\end{tabular}
\end{table} 

Following the prescription indicated in equation \ref{eq:PD}, we have calculated the relative values of the photo-detachment cross 
sections for the C$_2$N$^-$ anion. The data for the C$_2$H$^-$ anion we use here for comparison were presented before 
\cite{19GiGoMa.c2hm,20MaNoGo} and therefore we shall only discuss them during the analysis of the compared data below.

The two tables report three different choices for the exponential parameter already defined in the equation for the photo-detachment 
cross section. They were chosen according  to what is known about the experimental findings on the C$_2$N$^-$ anion as discussed, for 
example, by ref. \cite{09GeYaNe}. Briefly, 
the photo-detachment process originates from the linear anion in its ($X^3\Sigma^-$) electronic ground 
state with a molecular orbital (MO) configuration given as [core] $\pi^4$ $\sigma^2$ $\pi^2$. Removal of an electron from the highest 
occupied $\pi$ or $\sigma$ MO produces  the neutral  ($X^2\Pi$) or the neutral ($a^4\Sigma^-$) final states since they are both accessible
from the initial anionic states. As already shown in previous work \cite{09GeYaNe,07ZhGeNe},
photo-detachment from a $\pi$ or $\sigma$ MO at threshold 
energies can be mainly described as proceeding via either $s+d$ partial-wave contributions from the ejected electron or via a 
$p$-wave scattering contribution. Accordingly to the previous definition of the PD cross section, we can then assign to each of the outgoing 
electron channels different values of the exponential parameter $p = 0.5$ for s-wave scattering, 1.5 for $p$-wave scattering and 2.5 for 
$d$-wave scattering. The two tables therefore report our relative estimates of the cross sections for all three possible values of 
the exponential parameter. The data in Table \ref{tab:photo1} present our results for the case in which the laser frequency is 
chosen to be resonant with the value of the electron affinity (EA) for the present anion, while the data in Table  \ref{tab:photo2}
indicate the cross section values when 
the laser frequency  is increased by 50 cm$^{-1}$ above threshold. The data clearly indicate that the relative values of the 
PD cross sections are affected by both the initial value of the anion rotational state $N$ and the specific dominant partial-wave for 
the outgoing electron. Such difference will clearly reflect in the dynamical modelling of the loss rates in the trap that we will 
further present below.

\begin{figure}
\includegraphics[angle=270,width=0.47\textwidth]{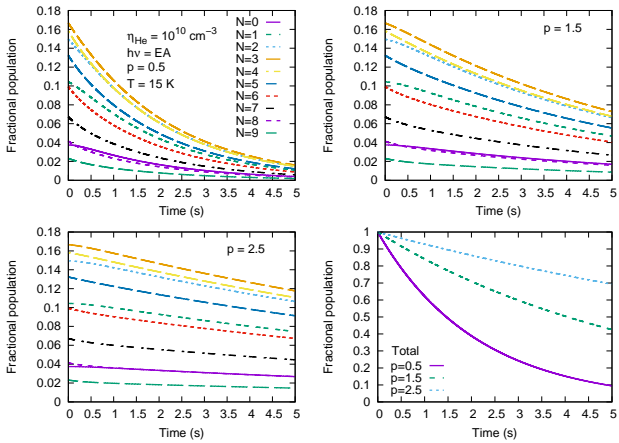} % Here is how to import EPS art
\caption{\label{fig:lossesB} Computed population evolution results for the PD process  for the case of the C$_2$N$^-$ anion. Different
operating conditions are shown in the  panels, while the lower right panel reports all the changes in the total  loss rates as the exponent 
$p$ is modified in estimating the PD cross sections used to generate the rates.  See main text for further details.}
\end{figure}

The computational results for the case of the C$_2$H$^-$ anion in the trap were already reported earlier by us in previous work 
\cite{19GiGoMa.c2hm,20MaNoGo} and will not be repeated here. New calculations have been carried out for comparison and are given for 
the C$_2$N$^-$ anion by the four panels shown in Figure ~\ref{fig:lossesB}. 
The laser action is switched on after 5 s, which is well beyond the time needed for the collisional equilibration of the populations of the
rotational levels to the temperature of the trap of 15 K. Earlier experiments \cite{09GeYaNe} indicate that two peaks are associated to the 
transitions to the neutral ($X^2\Pi$) state due to the splitting into two spin-orbit components. In the case of the C$_2$N that splitting was
found to be  38 cm$^{-1}$, close to earlier estimates and calculations of 40 cm$^{-1}$. We shall not include this value in estimating the
energy of the ejected electron since the EA value is much larger ( i.e. around 2.75 eV)  and the location of the upper ($a^4\Sigma^-$) is 
about 9,000 cm$^{-1}$ above the lower ($X^2\Pi$) neutral state.
The photo-detachment process into the latter neutral product will therefore involve at threshold the $p$ values of either 0.5 or 2.5. The data
presented by Figure ~\ref{fig:lossesB} in three of the panels show changes in trap operation for the case of $s$-wave scattering, while all
three options for the $p$ parameter are reported in the lower-right panel in that same figure to compare their effects on the total anion 
loss rates.
 
The most dramatic difference with the earlier studies we have done on the PD population evolutions in small molecular anions in cold traps
\cite{19GiGoMa.c2hm,18GiLaHe,20MaNoGo,20SiNoMi} is the far higher number of populated rotational 
states of C$_2$N$^-$ even at the low temperature of 15 K in the trap.
In the data shown in all panels the laser frequency has been kept resonant with the molecular EA of 2.7489 eV and the density of the buffer
gas kept constant at 10$^{10}$ cm$^{-3}$. The upper left panel employs the $p$ exponent at 0.5, a value corresponding to the expected
threshold behaviour of the ejected electron from the ground  ($X^3\Sigma^-$) state of the anion undergoing photo-detachment of its outer  
$\pi$ MO into a free electron at threshold energy and the ($X^2\Pi$) state of the neutral. We see that all rotational states up to $N=8$ 
 are populated after collisional equilibration of the molecular rotational state population 
within the trap. On the reported time scale, the PD laser is switched on at t=0  and the first five second evolution of fractional
population losses are shown. We clearly see that essentially a uniform exponential decay if followed separately  by each of the populated states in
the trap.
 
As a different $p$ exponent value is chosen, however, we see that the photo-detachment loss rates on the upper right panel correspond now to 
a threshold electron being ejected from an outer $\sigma$  orbital of the initial anion into a free electron plus the next ($a^4\Sigma^-$)
state of the final neutral molecule. Here the loss rates evolve much more slowly and on the same time scale we see that their relative
populations remain much higher in the trap. Such differences of behaviour should be amenable to detection in possible experiments. When 
the $p$ exponent is further changed to the value corresponding to the higher energy component in the partial-wave expansion of the ejected
electron from an outer  $\pi$ MO into the ($X^2\Pi$) ground electronic state of the neutral molecule, we see from the changes of fractional
populations of the rotational states populated initially at 15K (lower-left panel in the figure) that the decay rates become even smaller and
some of the higher rotational states are hardly depleted during the shown time interval that was essentially emptying them in the data of the
upper-left panel. This clearly indicates that the dominant partial-wave component for the ejected electron markedly affects the 
photo-detachment efficiency described by the value of the PD rates included in the previous set of coupled evolutionary equations.
 
A summary of such effects is presented in the lower-right panel of Figure \ref{fig:lossesB}, where the total values of the fractional
population losses are given for the three different choices of the $p$ exponent in the expression for the PD cross section shown earlier.
We clearly see marked differences in the rate losses within the allotted time interval, 
indicating again that such differences in trap depletion behaviour could  be experimentally detectable.
  
 \begin{figure}
\includegraphics[angle=270,width=0.47\textwidth]{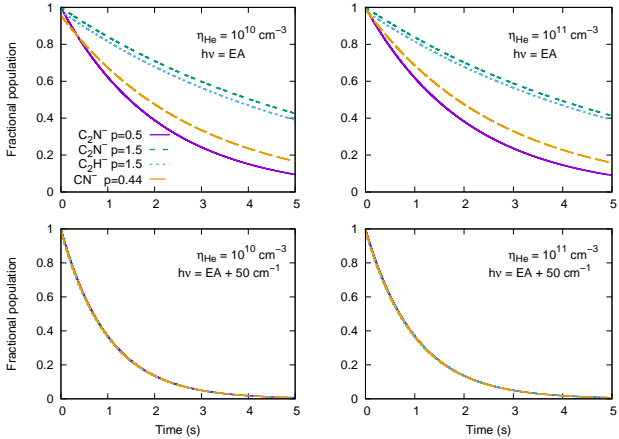} % Here is how to import EPS art
\caption{\label{fig:lossesC} A comparison of computed results for the PD-induced rotational population decays for three different polar 
anions reported in the captions, under different trap conditions. The trap temperature is set to 15 K, while the two top panels select 
the laser frequency to be resonant with each EA value of the anion. In the lower two panels the laser frequency is more energetic by 
50 cm$^{-1}$. See main text for further details.}
\end{figure}

Another interesting comparison is reported in the four panel of Figure ~\ref{fig:lossesC} where we compare the laser-induced population 
losses in the traps for two different values of the buffer gas density: on the two left panels that value is kept at   
10$^{10}$ cm$^{-3}$, while the data in the two left panels is taken to be 10$^{11}$ cm$^{-3}$. The laser frequencies are selected to be
resonant with each of the chosen molecules in the data reported by the two panels on the left of the figure, while the data in the two 
panels on the right show an increase of the laser photo-detachment energy by 50 cm$^{-1}$. Besides showing the variations of the computed
global losses when varying the operational conditions of the trap for the two title anions,  we have also added as a comparison the results 
we have obtained for another system studied in photo-detachment experiments in traps: the CN$^-$ anion which we have discussed in some of 
our recently submitted work \cite{20SiNoMi,20GoMaWe}. 
The latter anion is a closed-shell, ($^1 \Sigma$) molecule with a permanent dipole moment of 0.709 
Debye. Its EA value has been recently measured very accurately to be 3.6843(2) eV \cite{20SiNoMi} and at threshold the process is considered
to chiefly give rise to an s-wave scattering electron modified by the dipole of the anion, so that a value of the $p$ exponent of around 
0.44 can be realistically chosen, as discussed in an upcoming publication\cite{20SiNoMi}.

The upper-left panel of Figure ~\ref{fig:lossesC} presents two choices for the present C$_2$N$^-$ anion of the $p$ exponent describing its PD
cross section energy dependence and associated to dominant $s$ -wave and dominant $p$-wave behaviour of the ejected electron at threshold, as
already discussed earlier. For the C$_2$H$^-$ anion the threshold behaviour of the ejected electron is known to be chiefly controlled by 
$p$-wave scattering \cite{07ZhGeNe} and therefore the total loss rates reported in that panel employ the $p=1.5$ exponential value. 
It is also interesting to note that the permanent dipole
moments, involved in the evaluation of the transition moments defining the PD cross sections, show the largest value (3.09 Debye) for the 
C$_2$H$^-$ anion, followed by the C$_2$N$^-$ ( 2.18 Debye) and then by the CN$^-$ ( 0.709 Debye). However, these values  are not
dominating the PD process in the trap since collisional efficiency and density of rotational states in the three systems are also important
contributors. Thus, we see that the slowest anion losses in time occur for the two largest exponents describing the dominant partial waves of
the ejected electrons. On the other hand, the two cases where at threshold $s$-wave scattering dominates exhibit here faster anion losses in 
the trap. 

This relative behaviour is also maintained  by the data seen in the upper-right panel in Figure \ref{fig:lossesC}: 
the trap setting which has been
changed is the density of the buffer gas atoms, thus making the PD process to be driven by the increased collision numbers and the increased
role of the collisional repopulations of the rotational states of the trapped molecule, referred to earlier as a 
`collision-dominated'  photo-detachment process. We can therefore say that the more important differences between similar molecular 
anions are linked to their structural properties which decide the dominant partial-wave  during the threshold ejection of the extra electron.

As the photon energy is increased well above the EA thresholds in each system, as shown by the lower two panels of Figure \ref{fig:lossesC},
we see how remarkably the anion losses are driven by the energetics of the laser and not by the structural features of the individual target: 
the present examples now exhibit markedly higher rates of decaying fractional populations of their anions. Additionally, the rates are the
same for all of them without having their structural differences playing any significant role during the trap processes. We also clearly 
see that increasing the presence of collisional effects (in the lower-right panel) does not have any significant role on modifying the 
anion loss rates. In all the present systems we also note that a single exponential decay can describe their rates, indicating that 
no structural features involving relative re-population between different rotational states are important enough to modify the rates during
decay processes. This fact was also indicated by the individual decay rates shown by different rotational states of the C$_2$N$^-$ anion
shown in the panels of Figure \ref{fig:lossesB}. At the considered trap conditions of 
temperature much fewer 
rotational states ( i.e. up to $j=3$) are significantly populated for the C$_2$H$^-$ anion and for the
 CN$^-$ anion ( also up to $j=3$).

Finally, by changing the value of $\alpha$ in equation \ref{eq.alpha} the time for anion loss for PD is modified. This accounts for details
of a PD experiment as shown in our previous study of PD processes involving the ortho-NH$_2$$^-$ and the para-NH$_2$$^-$ anions
\cite{18GiLaHe} where the value of $\alpha$ was scaled to experimental ion losses. An order of magnitude decrease in the value of $\alpha$
leads to an order of magnitude increase in trap time. Thus, our calculations also find that for $\alpha$ values scaled down to 0.1 or 0.01
 the survival of anions in the traps can be extended up hundreds of seconds for the present systems.

\section{\label{sec:conclusions} Conclusions}

In this work we have analysed in detail the structural features of the interaction potential energy between He atoms and two small
linear anions which have been often considered to be present (albeit not yet detected) in ISM when modelling the 
chemistry of such environment, C$_2$H$^-$ ($X^1\Sigma^-$) and C$_2$N$^-$ ($X^3\Sigma^-$). We have employed the \textit{ab initio}
computed PESs to further evaluate the network of state-changing cross sections involving their lower rotational states in each of 
these systems which would be significantly populated at the expected temperatures of interstellar conditions. The computed cross 
sections were then employed to obtain the corresponding state-changing rates over a range of temperatures also significant for 
interstellar environments. The calculations found different relative sizes between rates in both systems and the preference of 
transitions involving the $\Delta N$=2 selection rule for C$_2$H$^-$ anion while transitions with $\Delta N$=1 are dominating 
for the C$_2$N$^-$ anion. These differences of behaviour can be linked to the structural differences between coupling dynamical terms 
in the interaction potentials which guide rotational state-changing collisions for either molecular anion. Differences in propensity 
rules involving  transitions with larger $\Delta N$ values also indicate that the C$_2$N$^-$ anion would favour transitions with odd 
values of $\Delta N$ while the even values are those favoured for C$_2$H$^-$. On the whole, however, our calculations find that 
both anions behave similarly in terms of efficiency of collisional state-changing probabilities while specific differences between 
processes are clearly present and can be amenable to observations.

We have also modelled the  behaviour of both molecular anions when uploaded to cold temperature ion traps (e.g. for temperatures around 
15 K as an example) and using He as a buffer gas in the traps. In particular we have modelled the photo-detachment processes of both 
anions after collisional equilibration of their populated rotational states at the trap temperature. By modifying trap operating conditions
such as the density of the buffer gas, energy of the photo-detaching laser and its power via  scaling the corresponding PD rates
driven by the laser, we have found differences of behaviour between the two systems and indicated their crucial role in controlling the global
loss rates of trapped anions of the dominant partial-wave of the ejected threshold electron. The latter parameter depends on the selection
rules of the PD process and on the initial and final electronic states of the anionic  and of the final neutral molecule. The increase of 
the photon energy of the laser is also significantly affecting the loss rates and is seen in our modelling that can shift operating 
conditions from being either chiefly collision-driven events or laser-driven events, as we have already discussed in our earlier 
work \cite{19GiGoMa.c2hm,20MaNoGo}.

The availability of accurately determined state-changing rotational rates at the temperature of the ISM for molecular ions employed in these
networks are expected to document more realistically the values for such quantities within kinetic models of chemical processes and energy
exchange processes in that environment. We also suggest that our specific results for estimating the rates of ion loss changes in cold ion
traps under different trap conditions could help in the preparation of future experiments on these anions.

\section{\label{sec:Supplementary Material} Supplementary Material}

There are 4 files in the Supplementary material: The  V$_\lambda$   multipolar coefficients  for
 C$_2$N$^-$/He PES, the computed rates for C$_2$H$^-$/He, the computed rates for C$_2$N$^-$/He and a $readme$ file explaining the format and units of the rates.

\begin{acknowledgments}

We are very grateful to professor M.L. Senent for having generously provided us with her computed potential energy values for the  
C$_2$H$^-$ system interacting with He atoms. F.A.G. and R.W. acknowledge the financial support by the Austrian Science Fund (FWF), 
Project P29558-N36 . 
One of us (J.F.) thanks  Europlanet
for a travel grant, the European COST action {\it Our Astro-Chemical History} for 
financing a short-term scientific mission (STSM) at the University of Innsbruck, 
and the computer center WCSS in Wroclaw for computational resources made
available under grant number KDM-408.

\end{acknowledgments}

\section{\label{data availability statement} Data Availability Statement}

The data that supports the findings of this study are available within the article and its supplementary material.

\bibliography{C2Nm}% Produces the bibliography via BibTeX.

\end{document}